# PLANNED SEARCH FOR LIGO/GBM COINCIDENCE IN THE LIGO O1 DATA RUN

Jordan Camp[1], Lindy Blackburn[2], Michael S. Briggs[3], Nelson Christensen[4],
Valerie Connaughton[3], Leo Singer[1], Peter Shawhan[5], John Veitch[6]

[1]*Astrophysics Science Division, Goddard Space Flight Center, Greenbelt, Md 20771*
[2]*Harvard-Smithsonian Center for Astrophysics, Cambridge, Mass 02138*
[3]*CSPAR, University of Alabama, Huntsville, Al 35899*
[4]*Physics Department, Carleton University, Northfield, Minn. 55057*
[5]*Physics Department, University of Maryland, College Park, Md 20742*
[6]*Physics Department, Cardiff University, Cardiff, UK*

In the fall of 2015 the first scientific observing run (O1) of the advanced LIGO detectors will be conducted. Based on the recent commissioning progress at the LIGO Hanford and Livingston sites, the gravitational wave detector range for a neutron star binary inspiral is expected to be of order 60 Mpc. We describe here our planning for an O1 search for coincidence between a LIGO gravitational wave detection and a gamma-ray signal from the Fermi Gamma-ray Burst Monitor. Such a coincidence would constitute measurement of an electromagnetic counterpart to a gravitational wave signal, with significant corresponding scientific benefits, including revealing the central engine powering the gamma-ray burst, enhanced confidence in the event as a genuine astrophysical detection, and a determination of the relative speed of the photon and graviton.

## 1. LIGO

The remaining years of this decade are likely to see a direct detection of a gravitational wave (GW). A passing gravitational wave produces a differential strain (ratio of change in distance to distance between two points at rest) in space-time along orthogonal directions transverse to the direction of propagation, and can be observed through the relative timing of the passage of light waves along these directions. Because gravity is a very weak force, the strain expected at the earth from the gravitational waves of even the strongest astrophysical sources is very small, of order $10^{-21}$.

The general design of the Laser Interferometer Gravitational Wave Observatory (LIGO) gravitational wave detector is shown in Fig. 1, and is based on the principle of laser interferometry[1]. A very high level of displacement sensitivity, of order $10^{-18}$ m rms, is achieved through careful attention to the frequency and amplitude stability of the laser light, the losses of the interferometer optics, the seismic isolation system which decouples the interferometer from motions of the earth, and the suspension system which stabilizes and positions the optics. Kilometer scale detectors are required to enable sufficient strain sensitivity for gravitational wave detection.

The LIGO detectors in Hanford, Washington (Fig. 2) and Livingston, Louisiana have recently undergone an upgrade[2] to enhance their range for observation of the GW from a neutron star binary inspiral. Compared to the initial LIGO configuration (disassembled in 2010), the range will ultimately be increased by a factor of 10, from 20 Mpc to 200 Mpc. At the full 200 Mpc sensitivity, the detection rate of NS binary inspirals will be of order 40/yr.[3]

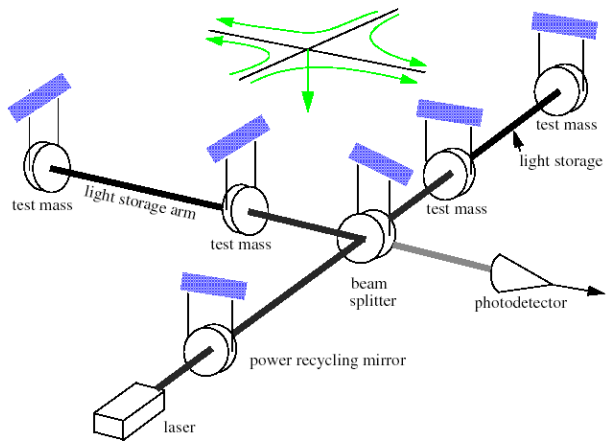 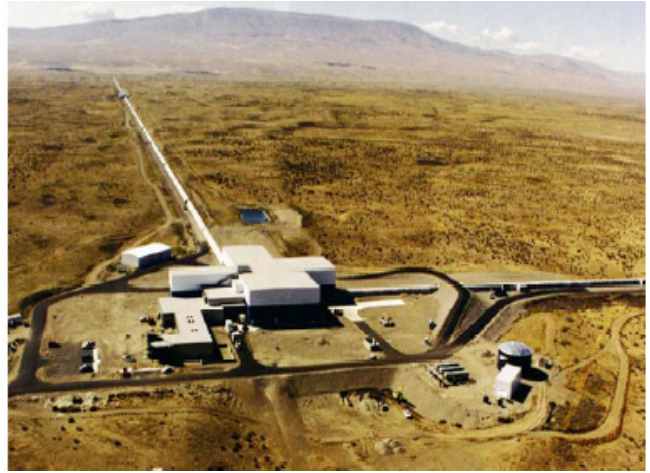

*Figure 1. Gravitational wave interferometric detector*     *Figure 2. Aerial view of LIGO site at Hanford, Washington*

At the time of this writing, the LIGO detector ranges were 35 Mpc and 69 Mpc for Hanford and Livingston, respectively. By the time of the first 3-month advanced LIGO observing run (O1, fall 2015), chances are good that both detectors will be operating in the range of ~60 Mpc. Subsequent planned runs include O2 (6 months, fall 2016, ~100 Mpc) and O3 (9 months, fall 2017, ~150 Mpc)[4].

## 2. Fermi GBM

The Fermi mission[5] (Fig. 3) was launched by NASA in 2008. Designed to survey the gamma-ray sky, it consists of two instruments: the imaging Large Area Telescope (LAT), designed to observe gamma-rays in the energy range of 20 MeV–300 GeV, and the Gamma-ray Burst Monitor (GBM)[6], sensitive to lower energy gamma-rays from 8 keV to 40 MeV.

GBM (Fig. 4) consists of 11 NaI detectors which have an energy range of up to 1 MeV, and 2 denser BGO detectors which have sensitivity up to 40 MeV. These detectors are affixed to the Fermi spacecraft in varying orientations, as shown in Figure 3. The NaI detectors have a cos θ response relative to angle of incidence, providing a position sensitivity to a gamma-ray source of ~5 degree by using the relative rates of differently-oriented detectors. GBM has a field of view (FoV) of 63% of the sky, ensuring that a gamma-ray source within the LIGO range that is beamed at the earth will have a good chance of detection.

## 3. Short Gamma-ray Bursts and GW Gamma-Ray Counterparts

The detection of a gamma-ray counterpart to a gravitational wave signal from an NS-NS or NS-BH merger will powerfully enhance the science of the compact merger event. In brief, the detection of a counterpart will: 1) test predictions of General Relativity, including the speed and polarizations of gravitational waves; 2) raise the confidence level of a GW detection by reducing false alarms; and 3) determine the engine of the sGRB (NS-NS or NS-BH) through the analysis of the GW.

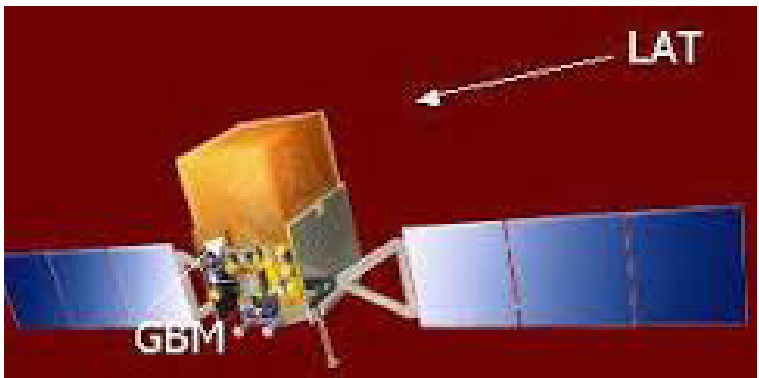

*Figure 3. Fermi spacecraft showing LAT and GBM instruments*

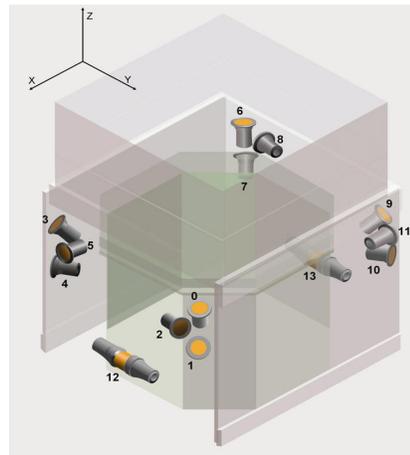

*Figure 4. GBM detector array. Detectors 1–11 are NaI, 12–13 are BGO*

The short gamma-ray burst[7] (sGRB) is an extremely energetic event powered by accretion onto a central compact object, which produces a jet of gamma-rays lasting less than two seconds. Strong circumstantial evidence identifies the sGRB engine as an NS-NS or NS-BH merger, whose inspiral is also the main anticipated source of gravitational waves for LIGO. The sGRB, observed with a rate density of ~10 Gpc$^{-3}$yr$^{-1}$ (ref. 8), is thus an excellent candidate for a GW/gamma-ray coincidence. Given an expected NS-NS horizon for the LIGO O1, O2, and O3 runs of 60 Mpc, 100 Mpc, and 150 Mpc, respectively, and including a range enhancement factor of ~2 from the gamma-ray coincidence[9], the number of GW/gamma-ray coincidences detected in these runs will be of order 0.02, 0.1, and 0.5 respectively. If the engine includes a BH (mass ~ 10 M$_\odot$), the horizon is doubled and the rates increase by a factor of 8.

A further interesting possibility for a gamma-ray counterpart is the production of a so-called sGRB precursor through the resonant cracking of a NS crust[10]. In this model, seconds before the NS merger of a binary inspiral, tidal deformation of a NS causes crust disruption by resonant excitation of a bulk-crust interface mode. The crust cracking then couples to the NS magnetic field to produce near-isotropic gamma-rays. Observation of a precursor in coincidence with a GW could determine the crust resonant frequency and provide information on the NS equation of state.

## 4. LIGO-GBM coincidence search: Detection Pipeline

The detection pipeline[11] for the LIGO-GBM coincident search is outlined in the flowchart of Figure 5. It consists of four steps: i) the identification of a GW trigger using a template-based analysis of the LIGO time-series data; ii) the extraction of the sky location of the GW event using a fast parameter estimation code BAYESTAR[12]; iii) the coherent analysis of the GBM detector data to determine consistency with the time and sky location of the LIGO trigger; and iv) the final output of the LIGO–GBM coincident events.

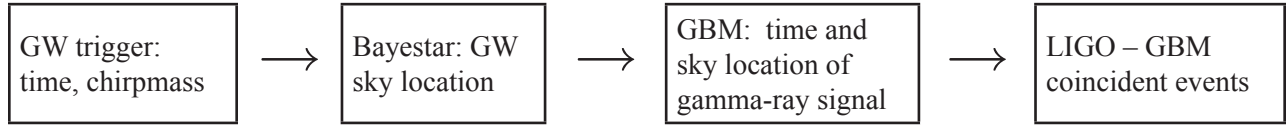

*Figure 5. Flowchart of LIGO-GBM analysis pipeline*

The initial step in the detection pipeline is the identification of a GW trigger. This is done primarily through the use of matched filtering, correlating the GW data in each detector against a bank of theoretical templates composed of model inspiral chirp signals with various component masses and possibly spin[13]. For the filter corresponding to a frequency domain signal $\tilde{h}(f)$, the matched filter produces an expected single-detector S/N of

$$\rho^2 = \int_0^\infty \frac{4|\tilde{h}(f)|^2}{S_n(f)} df \tag{1}$$

where $S_n(f)$ is the power spectral density of the stationary noise. When the event has a ρ value above a predetermined threshold (e.g., one false alarm per 10$^3$ yr), the event is considered a GW candidate.

Once a GW trigger has been generated, the algorithm BAYESTAR[12] is used for sky localization. BAYESTAR is a rapid Bayesian position reconstruction code that produces accurate sky maps less than a minute after a BNS merger detection, by using the times, phases, and amplitudes of the GW coincident signals. This allows the delivery of real-time alerts to the astronomical community.

The gamma-ray signal is determined using a coherent offline analysis of the 14 GBM detectors, allowing the possibility of sub-threshold detection[11] with respect to the on-board two-detector coincidence. This involves the maximum likelihood Λ$_{GBM}$(d, H), which is calculated as the probability of measuring the observed GBM data $d$ in the presence of a model signal $H$ relative to the probability from background fluctuations alone:

$$\Lambda_{GBM} \sim \prod_i \frac{\sigma_{n_i}}{\sigma_{d_i}} \exp\left(\frac{(d_i - n_i)^2}{2\sigma_{n_i}^2} - \frac{(d_i - n_i - r_i s)^2}{2\sigma_{d_i}^2}\right) \quad (2)$$

where for a given model (time, duration, spectrum, sky location) $d_i$ is the data and $n_i$ is the expected contribution from background; $\sigma_{d_i}$ is the expected standard deviation of the total measured counts and $\sigma_{n_i}$ is the standard deviation from background alone. The instrumental response $r_i$ is a function of source spectrum and location and when multiplied by source amplitude at the earth, $r_i s$ gives the expected contribution to each of the measured counts from the presence of the model signal.

The final probability of a GW / gamma-ray coincidence is then given by

$$\Lambda_{GW-GBM} = \int P_{GW}(\Omega) \Lambda_{GBM}(\Omega) d\Omega \quad (3)$$

where $P_{GW}(\Omega)$ is the GW probability distribution over the sky, derived from BAYESTAR.

The GW-triggered approach presented here will be able to provide an automated characterization of the high-energy sky during the time of known NS/NS mergers seen by Advanced LIGO. An alternate strategy to search for GW's associated with GRB's involves triggering a deep search in GW data using the time and sky location of known GRB's[14]. Finally, this LIGO-GBM search also complements related efforts with ground-based optical telescopes.[15]

## 5. Summary

We have presented our plan for the upcoming LIGO O1 science run, which involves a search for a GW / gamma-ray coincidence. At the expected O1 detector ranges in fall 2015, a detection of a coincidence is not likely. However the search will allow us to implement and test the detection pipeline, and be ready for the future O2 and O3 runs where a GW / gamma-ray coincidence becomes increasingly plausible.